\documentclass[smallextended]{svjour3}       
\newif\ifdraft \drafttrue
\smartqed  
\usepackage{graphicx}
\usepackage{natbib}
\bibpunct{(}{)}{;}{a}{}{,}
\usepackage{color}
\usepackage[linktocpage=true]{hyperref}
\hypersetup{
            colorlinks=true,
            linkcolor=black,
            filecolor=black,
            citecolor=black,
            urlcolor=blue
           }
\definecolor{Dblue}{rgb}{0.070,0.070,0.312}
\newcommand{\Blb}[1]{\textcolor{Dblue}{\bf #1}}
\newcommand{\web}[1]{\Blb{\url{#1}}}
\newcommand{\vex}{\vspace{1ex}}
\newcommand{\dint}{\int\hspace{-0.25em}\int}
\newcommand{\getlength}[1]{\ifx#1\end \let\next=\relax
            \else\advance\count255 by1 \let\next=\getlength\fi \next}
\newcommand{\ifnularg}[1]{ \count255=0 \getlength#1\end \ifnum\count255=0 }
\newcommand{\ifm}{\makebox{}\ifmmode}
\long\def\ifundefined#1#2#3{\expandafter\ifx\csname
  #1\endcsname\relax#2\else#3\fi}
\newcount\switch
\newcommand{\Begmat}{\ifm\switch=1\else\switch=0$\fi}
\newcommand{\Endmat}{\ifnum\switch=0$\fi}
\newcommand{\beq}   { \begin{eqnarray} }
\newcommand{\eeq}[1]{ \ifnularg{#1} \end{eanarray} \else
                      \label{#1}\end{eqnarray}    \fi }
\newcommand{\eeql}   { \end{eqnarray} }
\newcommand{\eeqn}   { \nonumber \end{eqnarray} }

\newcommand{\dss}{\displaystyle}

\newcommand{\Frac}[2]{\frac{\displaystyle\strut #1}{\displaystyle\strut #2} }

\newcommand{\der}[2] {\Begmat \frac{ \partial #1 }{ \partial #2 } \Endmat }

\newcommand{\upp}[1]{ {}^#1{\scriptstyle\kern-0.3em . \kern0.15em } }

\newcommand{\lp}{ \left(  }
\newcommand{\rp}{ \right) }

\ifundefined{sun}{}{}

\begin{document}

\author{Leonid Petrov}
\title{The International Mass Loading Service}
\institute{L. Petrov \at
              ADNET Systems, Inc, Falls Church, VA 22043, USA \\
              Tel.: +703-556-8757         \\
              \email{Leonid.Petrov@lpetrov.net}  \\
}
\date{}
\maketitle

\begin{abstract}
   \par\vspace{-20ex}\par
   
   The International Mass Loading Service computes four loadings: 
a)~atmospheric pressure loading; b) land water storage loading; c) oceanic 
tidal loading; and d) non-tidal oceanic loading. The service provides to 
users the mass loading time series in three forms: 1)~pre-computed time series 
for a list of 849 space geodesy stations; 2)~pre-computed time series on the
global $1^\circ \times 1^\circ$ grid; and 3)~on-demand Internet service for a list 
of stations and a time range specified by the user. The loading displacements
are provided for the time period from 1979.01.01 through present, updated 
on an hourly basis, and have latencies 8--20 hours. 
\par\vspace{-4ex}\par
\end{abstract}

\section{Introduction}

   Loading is a crustal deformation caused by a redistribution of air or water
mass. In particular, loading caused by the redistribution of air mass is 
called atmospheric pressure loading, redistribution of continental water mass 
in a form of snow cover, soil moisture, and ground water causes continental 
water storage loading (sometimes also referred to as ``hydrological loading''), 
redistribution of oceanic water mass causes ocean loading, which in turn is 
sub-divided into the ocean tidal and ocean non-tidal loadings. The loading 
deformation on average has the rms of 3~mm but may reach 60~mm ($M_2$ tidal 
ocean loading near British island). Calculation of mass loading is 
a computationally intensive procedure. For a case of ocean loading, 
the coefficients of displacement expansion can be computed once and forever. 
Computation of other loadings requires knowledge of surface pressure changes 
caused by a redistribution of air and water masses that is highly volatile and 
cannot be computed beforehand. Acquiring information about time series of 
global mass redistribution poses a serious logistical problem that impeded 
implementation of data reduction for mass loading.

  Realizing the magnitude of logistical problem, the mass loading service 
at NASA Goddard Space Flight Center was established in December 2002 
\citep{r:aplo}. The service was limited to the atmospheric pressure loading. 
It provided for the geodetic community the time series of 3D displacements 
of several hundred space geodesy stations, as well as the global displacement 
field at the $1^\circ \times 1^\circ$ grid. The time series were updated daily
and had latency around 3 days. The service quickly became very popular and 
became the main source of information about atmospheric pressure loading.

  Recently, a decision was made to make a deep upgrade of the service.
The upgrade targeted the following areas:
a)~to extend the service to loading caused by land water storage and
non-tidal ocean loading; b)~to support high resolution models of the 
atmosphere and land water storage; c)~to take into account effects of local 
topography on surface pressure in mountain regions; d)~to improve latency;
e)~to provide a user an ability to compute loading for user-selected
stations on-demand. In the rest of the paper I will describe the approaches 
used for this upgrade.

\section{The use of high resolution models for loading computation}

  The original atmospheric pressure loading service used the 2D NCEP 
Reanalysis surface pressure field \citep{r:ncep} at a regular grid 
with a spatial resolution $2.5^\circ \times 2.5^\circ$. Modern models 
have much higher resolutions: for instance, the GEOS-FP model has resolution 
$0.3125^\circ \times 0.25^\circ$. The traditional approach for loading 
computation at a point with coordinate $\vec{r}$ involved a numerical 
evaluation of the integral of a convolution type \citep{r:far72}:
\beq 
  \begin{array}{lcl}
     \vec{u}_r(\vec{r},t) & =  & \dss\dint\limits_{\!\!\!\!\!\!\Omega} L(\phi',\lambda') \, 
                                 \Delta P(\vec{r}\,',t) \, 
                               G_{\mbox{\sc r}}(\psi) \cos \phi' d \lambda' d \phi' 
     \vex \\
     \vec{u}_h(\vec{r},t) & = & \dss\dint\limits_{\!\!\!\!\!\!\Omega} \vec{q}(\vec{r},\vec{r}\,') \, 
               L(\phi,\lambda)\, \Delta P(\vec{r}\,',t) \, 
               G_{\mbox{\sc h}}(\psi) \cos \phi' d \lambda' d \phi' ,
  \end{array}
\eeq{e:e1}
where $\Delta P(\vec{r}\,',t)$ is the pressure caused by mass redistribution,
$L(\phi,\lambda)$ --- is the land-sea mask, the share of land in 
an elementary cell, and $G(\psi)$ are the Green's functions defined as
\beq
  \begin{array}{lcl @{\qquad\quad} lcl }
     \dss
     G_{\mbox{\sc r}}(\psi) & = & \Frac{f a}{g_0^2}  \dss\sum_{n=0}^{+\infty} 
                                  h^{'}_n P_n (\cos \psi) 
     &
     G_{\mbox{\sc h}}(\psi) & = & -\Frac{f a}{g_0^2} \dss\sum_{n=1}^{+\infty} 
                                  l^{'}_n \frac{\partial 
                                  P_n (\cos \psi)}{\partial \psi}
  \end{array}.
\eeq{e:e2}

  The problem is that this algorithm has complexity $O(d^4)$, where $d$ is the 
spatial grid size, i.e. it grows very rapidly with an increase of spatial 
resolution. It becomes impractical to use convolution for loading computation 
using models with a high spatial resolution. The alternative is to use 
the spherical harmonic transform approach. The algorithm involves 
the following steps:

\begin{enumerate}
    \item forming the pressure difference with respect to the average;

    \item transforming the surface pressure field to the regular grid 
          with a higher resolution (upgridding): $2(D+1)+1 \; \times \; 4(D+1)$ 
          over latitude and longitude, where $D$ is degree of the expansion;

    \item multiplying the surface pressure field with the land-sea mask 
          defined as a share of land in a cell;

    \item spherical harmonic transform of degree/order D;

    \item scaling the output of the spherical harmonic transform with Love 
          numbers $h'_n$ and $l'_n$ of the corresponding degree $n$:
\beq
    \begin{array}{lcl}
        V^m_n(t) & = & \Frac{1}{\bar{\rho}_\oplus \, g_0} \Frac{3 h'_n}{2n+1} 
                    \dss\dint\limits_{\!\!\!\!\!\!\Omega} 
                    L(\phi,\lambda) \, \Delta P(t,\phi,\lambda) \, 
                    Y^m_n(\phi,\lambda)
                    \cos \phi \, d\phi \, d\lambda \vex \\
        H^m_n(t) & = & \Frac{1}{\bar{\rho}_\oplus \, g_0} \Frac{3 l'_n}{2n+1} 
                    \dss\dint\limits_{\!\!\!\!\!\!\Omega} 
                    L(\phi,\lambda) \, \Delta P(t,\phi,\lambda) \, 
                    Y^m_n(\phi,\lambda)
                    \cos \phi \, d\phi \, d\lambda \\
    \end{array},
\eeq{e:e5}
     where $\bar{\rho}_\oplus$ is the mean Earth's density and $g_0$ is the 
equatorial gravity acceleration. The expression under the integral is 
the spherical harmonics $\lp\rp{}^m_n$ of the the pressure field with 
the land-sea mask applied.

     \item inverse spherical harmonic transform:

\par\vspace{-3ex}\par
\beq
    \begin{array}{lcl}
       D_U(\phi,\lambda) & = & \dss\sum\limits_{i=0}^{i=m} \sum\limits_{j=-n}^{j=n} V^i_j\, 
                                        {Y^i_j}^*(\phi,\lambda) \vex  \\
       D_E(\phi,\lambda) & = & \dss\sum\limits_{i=0}^{i=m} \sum\limits_{j=-n}^{j=n} H^i_j\, 
                                        \der{{Y^i_j}^*(\phi,\lambda)}{\lambda} \vex \\
       D_N(\phi,\lambda) & = & \dss\sum\limits_{i=0}^{i=m} \sum\limits_{j=-n}^{j=n} H^i_j\, 
                                        \der{{Y^i_j}^*(\phi,\lambda)}{\phi}  \vex  \\
    \end{array}.
\eeq{e:e6}

\end{enumerate}

  This algorithm is equivalent to eqn~(\ref{e:e1}) when 
$D \longrightarrow \infty$, but it has complexity $O(d^3)$. It outperforms the
convolution algorithm when $D > 30$. Numerical tests showed that in order 
to have errors in loading computation everywhere on the Earth less than 
0.15~mm, degree/order 1023 is sufficient. It may sound counter-intuitive
why such high resolution ($0.088^\circ$) is needed, since the resolution
of numerical models is one order of magnitude coarser. We should bear in
mind that although the output of numerical models does not have signal at 
degree/order greater than 200--400, the product of the surface pressure and 
the land-sea mask is not band-limited and its spherical harmonic transform 
is not zero at any degree/order.

\par\vspace{-1ex}\par
\section{Mass redistribution models}

  Three numerical weather models developed at the NASA Global Modeling and 
Assimilation Office (GMAO) are used for loading computation:

\begin{itemize}
  \item MERRA (Modern-Era Retrospective analysis for Research and 
        Applications) \citep{r:merra}.
        Resolution: $0.67^\circ \times 0.5^\circ \times 72 \;\mbox{layers} 
        \times 6^h $, runs from 1979.01.01 through present, latency 
        $20^d$--$60^d$. This model is frozen and it is considered the most 
        stable.

  \item GEOS-FP (Global Earth Observing System Forward Processing) 
        \citep{r:geos}. 
        Resolution: $0.3125 ^\circ \times 0.25^\circ \times 72\; \mbox{layers} 
        \times 3^h $, runs from 2011.09.01 through present, latency 
        $6^h$--$15^h$. This is the operational model, updated approximately
        once a year.

  \item GEOS-FPIT (Global Earth Observing System Forward Processing 
        Instrumental Team) \citep{r:geos08}. 
        Resolution: $0.625 ^\circ \times 0.5^\circ \times 72\; \mbox{layers} 
        \times 3^h $, runs from 2000.01.01 through present, latency 
        $6^h$--$25^h$. In terms in stability this model is intermediate between
        MERRA and GEOS-FP, but it has a low latency.
\end{itemize}

   The surface pressure is computed from a 3D model. This process involves
several steps. Firstly, each column of the output at the native, 
{\it irregular}, terrain-following grid is interpolated to the column at 
a new regular grid that is formally extrapolated down to -1000~m and up 
to 90,000~m. Then the atmospheric pressure at a given epoch is expanded 
into the tensor product of B-splines over the entire Earth. Using the 
expansion coefficients, the pressure on the surface at resolution D1023 
($0.088^\circ \times 0.088^\circ$) is computed. The height of the surface 
is derived from $30'' \times 30''$ GTOPO30 model%
\footnote{\web{https://lta.cr.usgs.gov/GTOPO30}} by averaging over cells 
of the D1023 grid. Using the expansion coefficients, the atmospheric 
pressure on that surface is computed. This procedure mitigates effects 
of orthography: in mountainous regions a node of a coarse grid may fall 
into a valley or a ridge and therefore, may not be representative for 
an average pressure of the cell.

  Three land water storage models are used for loading computation:

\begin{itemize}
  \item GLDAS NOAH025 (Global Land Data Assimilation System) \citep{r:gldas}.
        Resolution: $0.25 ^\circ \times 0.25^\circ \times 3^h $, runs from 
        2000.01.24 through present, latency $35^d$--$70^d$. 

  \item MERRA TWLAND \citep{r:twland}.
        Resolution: $0.67^\circ \times 0.5^\circ \times 6^h $, runs from 
        1979.01.01 through present, latency $35^d$--$60^d$. This model is 
        considered the most stable.

  \item GEOS-FPIT TWLAND. Resolution: $0.625 ^\circ \times 0.5^\circ 
        \times 1^h $, runs from 2000.01.01 through present, latency 
        $6^h$--$25^h$. It was found that hourly time resolution is 
        excessive for loading computation. The resolution was reduced to 
        3~hours.
\end{itemize}

  Upgridding involves refining the pressure field according to the fine 
land-sea mask. If a cell at the new $0.088^\circ \times 0.088^\circ$ grid 
falls in the area that was ocean in the old grid, the pressure of the water 
equivalent of soil moisture and/or snow cover is computed by interpolation 
from surrounding cells that are land in the original grid with 
applying Gaussian smoothing.

  Non-tidal ocean loading is computed from the Ocean Model for Circulation 
and Tides (OMCT) \citep{r:omct}. The original resolution of the model 
is $1^\circ \times 1^\circ \times 6^h$, latency: $10^d$--$60^d$. However, 
the output of the original model is not available, only its spherical 
harmonic transform truncated at degree/order 100. Ugridding the OMCT model 
involves an iterative procedure that resembles the CLEAN algorithm used 
in radio astronomy for image restoration: it exploits the facts that the 
ocean bottom pressure is zero at land and the bottom pressure is relatively 
smooth in the ocean, except a jump to the zero at the shore.

  Two models of ocean tidal loading are used: the GOT4.8 \citep{r:got48}
and FES2012 \citep{r:fes2012}. They are upgridded to degree/order 2047 in 
a similar way as it was done for land water storage, except reversal of 
land and sea cells.

\par\vspace{-2ex}\par
\section{Processing pipeline}

  The two servers of the international mass loading service that work 
independently check every hour whether new data appeared. If the new 
data appeared, they
are downloaded, decoded, up-gridded, and the surface pressure anomaly
at the D1023 grid is computed by subtracting a model that includes the 
mean surface pressure value, sine and cosine amplitudes of pressure 
variations in a range of frequencies in the diurnal, semi-diurnal,
ter-diurnal and four-diurnal bands. Then the spherical harmonic 
transform of degree/order 1023 of the pressure field anomaly is 
computed and scaled by Love numbers of the corresponding order. The 
coefficients $V^m_n$ and $H^m_n$ in eqn (\ref{e:e5}) are stored.
They are used for loading computations in three ways:

\begin{enumerate}
   \item Computing loading at the D89 grid ($1^\circ \times 1^\circ$). 
         This is done in the following way: the spherical harmonic transform 
         of degree/order D1023 is padded with zeroes to degree/order D1079. 
         The coefficients $V^m_n, H^m_n$ are underwent the inverse spherical
         harmonic transform and produce the loading field in local Up, East, 
         and North direction at the D1079 grid ($1/12^\circ \times 
         1/12^\circ$). Every 12th element of the intermediate D1079 grid
         is written in the output file.

   \item Computing loading for a set of 849 commonly used 
         GNSS, SLR, DORIS, and VLBI stations. 

   \item Computing loading on-demand for the set of stations supplied
         by the user. A user fills the Web form where he or she specifies
         the model, the range of dates and the list of stations with
         their Cartesian coordinates. When the loading computation 
         is finished, a user can retrieve the files with results.
\end{enumerate}
  
  The loading displacements are computed using the Love numbers defined 
in the coordinate system with the origin at the center of mass of the 
total Earth: the solid Earth and the fluid under consideration. For some 
applications displacements with respect to the center of mass of the 
solid Earth are desirable. The loading Love numbers differ between these 
systems only for degree~1. The International Mass Loading Service computes
the differential loading displacements between these two systems. Such 
a displacement, called the ``degree one displacement'' uses only two 
terms of degree 1 in expression~(\ref{e:e6}): 
$\tilde{V}^m_1 = \Frac{1+h'_1}{h'_1} \, V^m_1$ and
$\tilde{H}^m_1 = \Frac{1+l'_1}{l'_1} \, H^m_1$.
When this degree one displacement is added to the displacement with
respect to the center of the total mass, the sum is the displacement
with respect to the center of mass of the solid Earth.

\section{Validation}

   VLBI observations for the period of 2001.01.01 -- 2014.07.01 were 
used for loading validation. The same technique was applied
as we used for loading validation in \citet{r:aplo}: the global
admittance factors were estimated from the data together with
estimation of site positions, velocities, the Earth orientation
parameters, source coordinates and nuisance parameters such as
clock functions and atmosphere path delays in zenith direction
(see Table~\ref{t:valid}). The partial derivative for admittance 
factors was the contribution of the loading displacement into path 
delay. If the model is perfect, the admittance factor will approach 
to unity.

\begin{table}
  \caption{Estimates of admittance factors for Up (UP),
           East (EA) and North (NO) components for three
           different loading models from the global least
           squares solution using geodetic VLBI group delays.}
  \begin{center}
    \begin{tabular}{lrcr}
         Atm GEOS-FPIT UP &  0.963 &$\pm$& 0.023  \\
         Atm GEOS-FPIT EA &  0.609 &$\pm$& 0.049 \\
         Atm GEOS-FPIT NO &  1.027 &$\pm$& 0.041 \\
                          &        &     &        \\
         Lws GEOS-FPIT UP &  0.955 &$\pm$& 0.016 \\
         Lws GEOS-FPIT EA &  0.804 &$\pm$& 0.029 \\
         Lws GEOS-FPIT NO &  0.886 &$\pm$& 0.024 \\
                          &        &     &        \\
         Lws NOAH025   UP &  1.220 &$\pm$& 0.013 \\
         Lws NOAH025   EA &  0.660 &$\pm$& 0.030 \\
         Lws NOAH025   NO &  0.826 &$\pm$& 0.033 \\
     \end{tabular}
  \end{center}
  \label{t:valid}
  \par\vspace{-4ex}\par
\end{table}

Surprisingly, the GEOS-FPIT land water storage model has admittance factors 
closer to unity than the GLDAS NOAH025 model. This is important for practical 
applications, since the GEOSFPIT model has much lower latencies than the GLDAS 
NOAH025 model.

\par\vspace{-2ex}\par
\section{Using the International Mass Loading Service}

\begin{figure}
   \caption{Mass loading caused by the $M_2$ ocean tide near Newfoundland
            island computed with two resolutions: 
            $0.01^\circ \times 0.01^\circ$ grid ({\it Left}) and
            $1.0^\circ \times 1.0^\circ$ grid ({\it Right})
           }
   \begin{center}
      \ifdraft
           \includegraphics[width=0.48\textwidth]{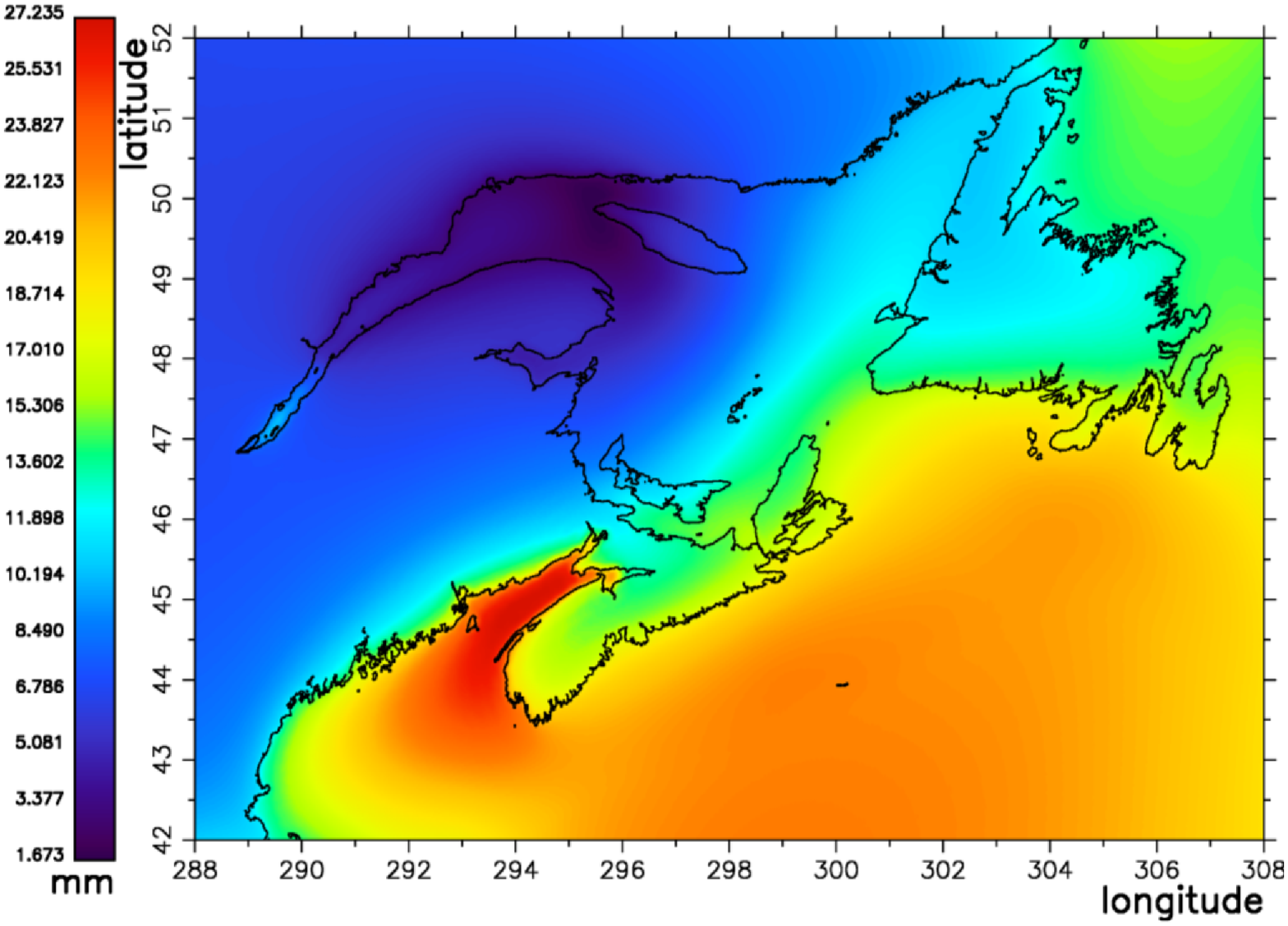}
         \else
           \includegraphics[width=0.48\textwidth]{m2_tide_01.eps}
      \fi
      \hspace{0.02\textwidth}
      \ifdraft
          \includegraphics[width=0.48\textwidth]{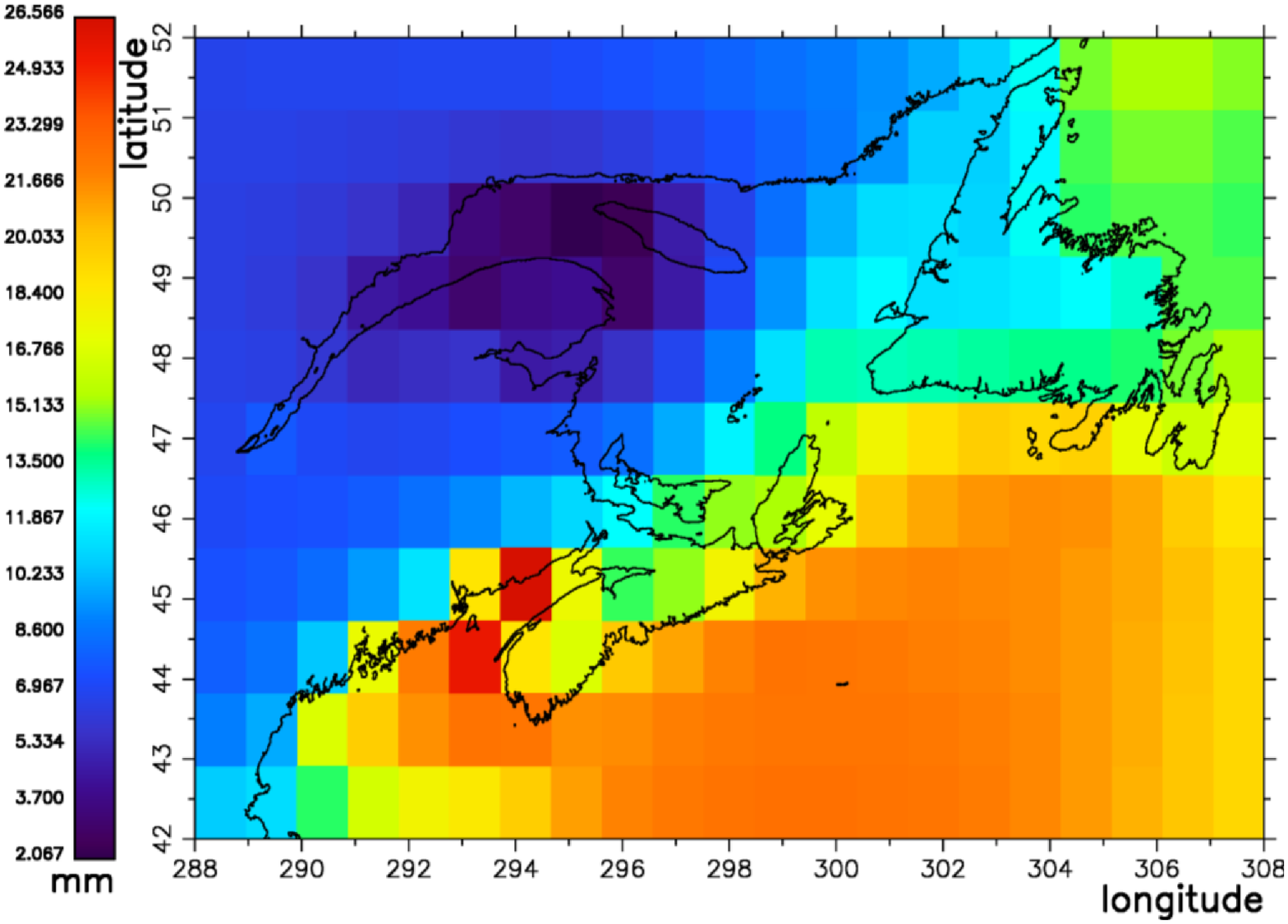}
        \else
          \includegraphics[width=0.48\textwidth]{m2_tide_03.eps}
      \fi
   \end{center}
   \par\vspace{-5ex}\par
   \label{f:m2_dspl}
\end{figure}

\begin{figure}
   \caption{The difference of mass loading caused by the $M_2$ ocean 
            tide computed with two resolutions: 
            $1.0^\circ \times 1.0^\circ$ grid versus 
            $0.01^\circ \times 0.01^\circ$ grid ({\it Left}) and
            $0.25^\circ \times 0.25^\circ$ grid versus 
            $0.01^\circ \times 0.01^\circ$ grid ({\it Left}) and
           }
   \begin{center}
      \ifdraft
          \includegraphics[width=0.48\textwidth]{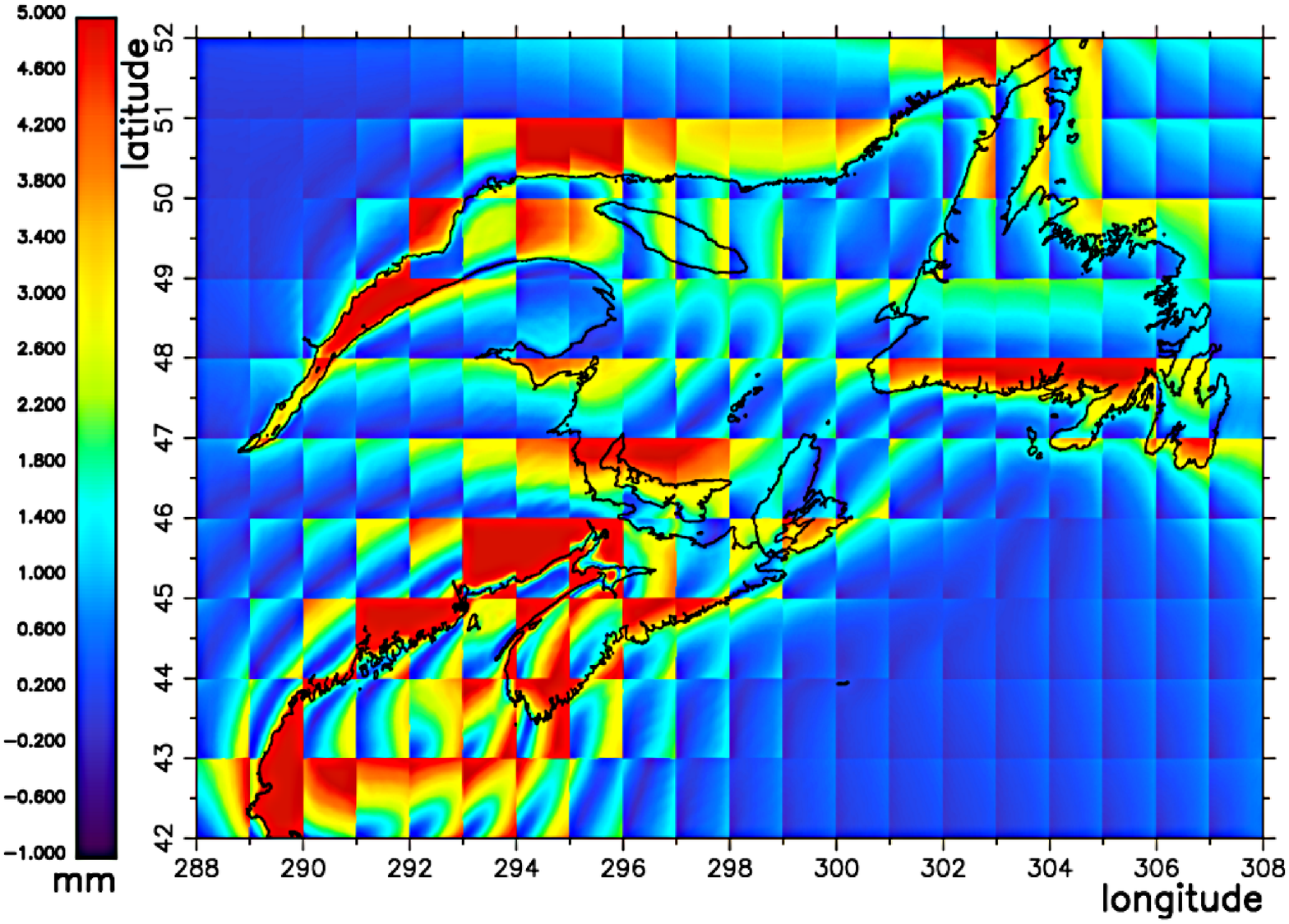}
        \else
          \includegraphics[width=0.48\textwidth]{m2_diff_01.eps}
      \fi
      \hspace{0.02\textwidth}
      \ifdraft
          \includegraphics[width=0.48\textwidth]{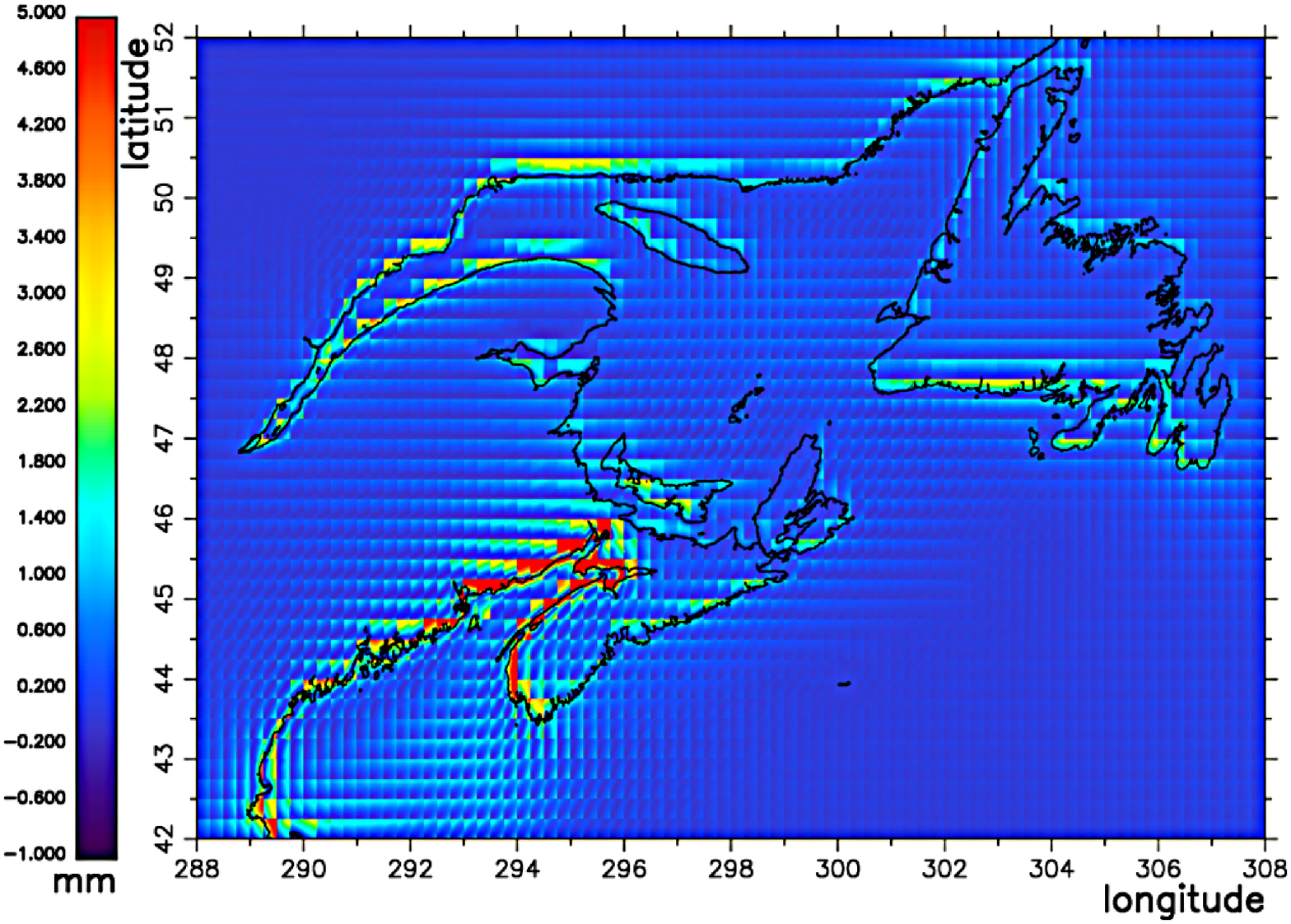}
        \else
          \includegraphics[width=0.48\textwidth]{m2_diff_02.eps}
      \fi
   \end{center}
   \par\vspace{-5ex}\par
   \label{f:m2_diff}
\end{figure}

   The gridded loading displacements are useful for visualization of the 
loading field and for computation of integrals over the area. However, a user 
should be aware that the field of loading displacement near the coastal area 
is not smooth. Therefore, using gridded loading for data reduction by 
interpolation the displacement field to the position of a given station 
may cause significant errors. This problem is illustrated in 
Figures~\ref{f:m2_dspl}--\ref{f:m2_diff} for a case of ocean loading near 
Newfoundland. The $M_2$ ocean loading displacement has the vertical amplitude 
$\sim\! 30$~mm, but interpolation errors exceed 30\% within 100~km of the 
coastal area when the $1.0^\circ \times 1.0^\circ$ grid is used. The errors 
are in excess of 30\% within 30~km from the coast when the 
$0.25^\circ \times 0.25^\circ$ grid is used. They fall below 1~mm only when 
the grid with a resolution $0.05^\circ \times 0.05^\circ$ or finer is used.

\begin{figure}
   \caption{User interface to the on-demand computation of mass loading
            displacements}
   \par\vspace{2ex}\par
   \begin{center}
        \includegraphics[width=0.75\textwidth]{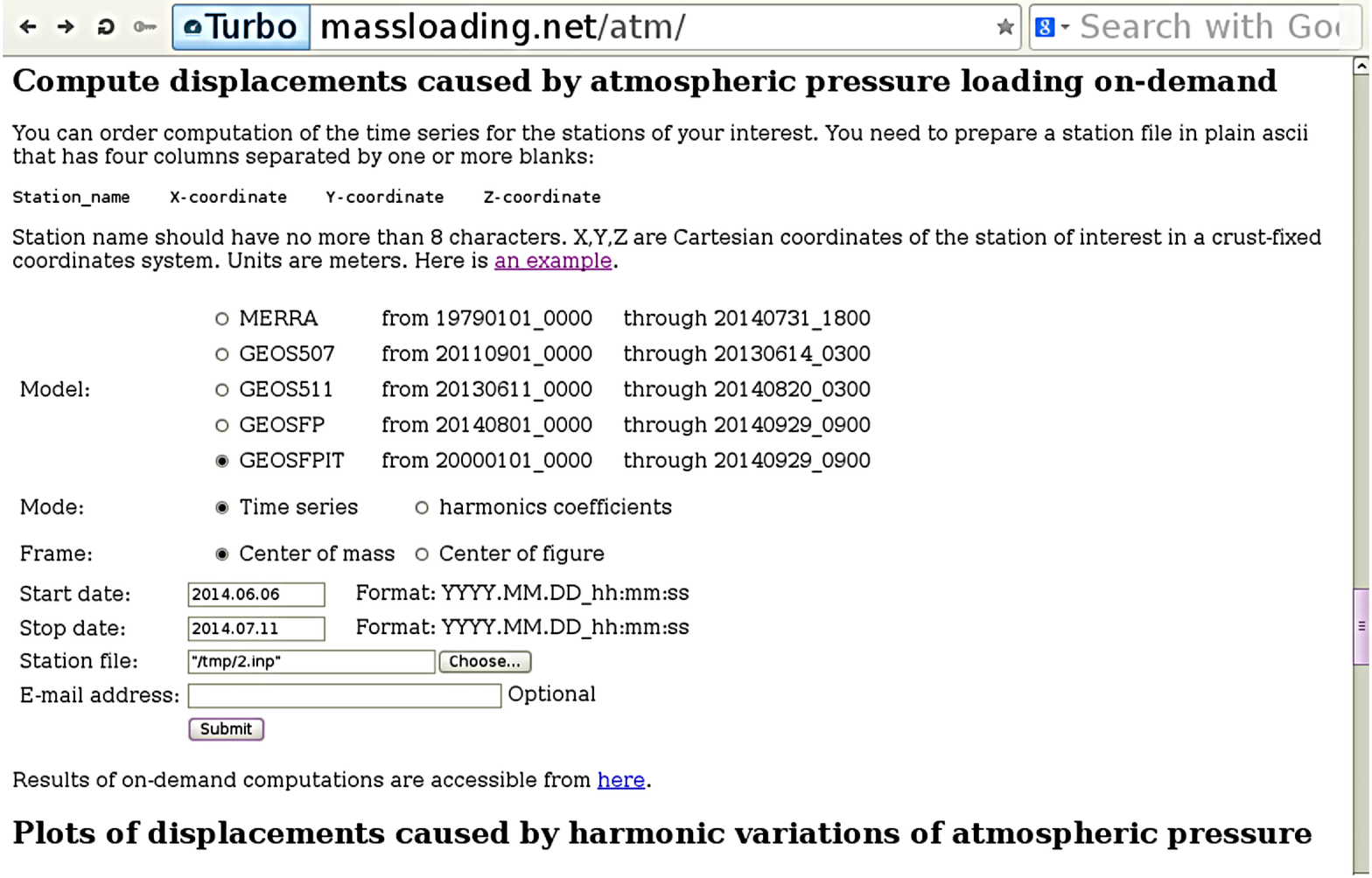}
   \end{center}
   \par\vspace{-6ex}\par
   \label{f:ond}
\end{figure}

  Gridded loading at $1^\circ \times 1^\circ$ or 
$0.25^\circ \times 0.25^\circ$ resolutions \underline{\bf should never be}
\underline{\bf used} for data reduction. The International Mass Loading 
Service computes loadings for 849 stations directly without the use of 
interpolation. This is sufficient for processing SLR, DORIS, and VLBI 
observations, since new stations are introduced infrequently. New GNSS 
stations are introduced much more frequently, and the situation when 
the GNSS station of interest is absent from the list of stations with 
pre-computed loading is more common. Figure~\ref{f:ond} shows the Web 
interface that implements the on-demand loading computation. 
The on-demand computational procedure uses $V^m_n(t)$ and $H^m_n(t)$ 
coefficients that are evaluated and stored as soon as new data arrive. 
For this reason, the on-demand procedure is relatively quick: 
computation of loading displacements for 20 stations for a 1~year 
interval with the time step of 3~hours, in total 58,400 loading 
displacements, takes 20~minutes.

\par\vspace{-2ex}\par
\section{Conclusions and future work}

  At present, the International Mass Loading Service offers to the geodetic
community computation of 3D displacements caused by the atmospheric pressure
loading, land water storage loading, tidal and non-tidal ocean loading,
free of charge, 24/7 with a latency from 8 hours (atmospheric and land water
storage loading) to 30 days (non-tidal loading). The URL of the primary
server is \web{http://massloading.net}, the URL of the secondary server
is \web{http://alt.massloading.net}. The loading displacement were validated
against the dataset of global VLBI observations for 2001--2014.

  Further development: a)~using weather forecast to 0--$24^h$ in the future. 
Latency will be eliminated. Accuracy degradation with respect to an assimilation
model: 20\% for the current 
instant; b)~using OPeNDAP protocol for data distribution; c)~automation of 
loading displacement ingestion: development of a client library that 
communicates with the loading servers automatically; d)~generation of time 
series of the sum of all loadings on the fly with a user-requested time 
step; e)~computing loading due to water level changes in lakes and 
big rivers.

This project was supported by NASA Earth Surface and Interior program, 
grant NNX12AQ29G.

\par\vspace{-4ex}\par


\begin{thebibliography}{11}

   \bibitem[Carrere et al.(2012)]{r:fes2012}
     Carrere~L, Lyard~F, Cancet~M, Guillot~A, Roblou~L (2012)
        FES2012: A new global tidal model taking taking advantage 
        of nearly 20 years of altimetry, Proceedings of meeting 
        ``20 Years of Altimetry'', Venice.

   \bibitem[Farrell(1972)]{r:far72}
     Farrell,~WE (1972) 
      Deformation of the Earth by Surface Loads, 
        Rev.\ Geophys.\ and Spac.\ Phys., vol.~10(3), 751--797

   \bibitem[Kalnay et al.(1996)]{r:ncep}
      Kalnay~EM et al. (1996) Bull. Amer. Meteorol. Soc., 77, 437--471


   \bibitem[Molod et al.(2012)]{r:geos}
      Molod~A, Takacs~L, Suarez~M, Bacmeister~J, Song~I-S, 
        Eichmann~A (2012) 
         The GEOS-5 Atmospheric General Circulation Model: Mean Climate and 
         Development from MERRA to Fortuna, 
        NASA/TM--2012, 104606, vol.~28 

   \bibitem[Petrov and Boy(2004)]{r:aplo}
      Petrov~L, Boy~J-P (2004)
      Study of the atmospheric pressure loading signal in VLBI observations,
     JGR, 10.1029/2003JB002500, vol.~109, No. B03405.

   \bibitem[Ray(2013)]{r:got48}
     Ray~R (2013) Precise comparisons of bottom-pressure and altimetric 
       ocean tides, JGR, 118, 4570--4584, doi:10.1002/jgrc.20336

  \bibitem[Reichle et al.(2011)]{r:twland}
    Reichle~RH, Koster~RD, De Lannoy~GJM, Forman~BA, Liu~Q, 
      Mahanama~SPP, Tour\'e~A (2011) J. Climate, 24, 6322--6338

  \bibitem[Rienecker, et al.(2008)]{r:geos08}
    Rienecker~MM, et al. (2008) The GEOS Data Assimilation System --- 
      Documentation of Versions 5.0.1, 5.1.0, and 5.2.0, 
      NASA/TM--2008--104606

  \bibitem[Rienecker, et al.(2011)]{r:merra}
    Rienecker~MM, et al. (2011) J. Climate, 24, 3624--3648

   \bibitem[Rodell et~al.(2004)]{r:gldas}
      Rodell~MP, et al (2004) The Global Land Data Assimilation System,
        Bull. Amer. Meteor. Soc., 85(3): 381--394

  \bibitem[Thomas(2002)]{r:omct}
     Thomas~M (2002) Ozeanisch induzierte Erdrotationsschwankungen,
     PhD 

\end{thebibliography}
\end{document}